\documentclass[12pt]{article}

\usepackage[left=2.5cm,top=3.5cm,right=2.5cm,bottom=2cm,nohead]{geometry}

\usepackage{setspace}

\usepackage{url}
\usepackage{multirow}
\usepackage{multicol}
\usepackage{graphicx}
\usepackage{caption}
\usepackage{subcaption}
\usepackage[export]{adjustbox}
\usepackage{color}

\usepackage{booktabs}

\usepackage{framed}

\usepackage{tcolorbox}

\newcommand{\sidebarcmd}{
\begin{center}
\begin{tcolorbox}[width=6.2in,
                  boxsep=10pt,
                  left=10pt,
                  right=10pt,
                  top=2pt,
                  ]
\begin{center}
\noindent{\bf Related Work on Software Popularity}\\
\end{center}
Yuan et al.~investigate 28 factors along eight dimensions to understand how high-rated Android applications are different from low-rated ones~\cite{Tian2015}. Their result shows that external factors, like number of promotional images and target SDK, are the most influential factors. Weber and Luo attempt to differentiate popular and unpopular Python projects on GitHub using machine learning techniques~\cite{Weber}. They found that in-code features are more important than author metadata features. Zho at al.~study the frequency of folders used by 140 thousands GitHub projects and the results suggest that the use of conventional folders may have an impact on project code popularity~\cite{Zhu}. By analyzing usage of Java APIs, Moleva states that the success of APIs are related to their usage trends~\cite{Mileva12}. Bissyande et al.~analyze the popularity, interoperability, and impact of various programming languages, using a dataset of 100K open source software projects~\cite{Bissyande2013}. Aggarwal et al.~study the effect of social interactions on GitHub projects' documentation~\cite{Aggarwal2014}. They conclude that  popular projects tend to attract more documentation collaborators. Finally, Figueiredo et al.~characterize the growth patterns of video popularity on YouTube~\cite{Figueiredo:2011}. The results shows that videos in the top lists tend to experience sudden significant bursts of popularity. To the best of our knowledge, we are first to track popularity over time on social code sharing sites, like GitHub. 
\end{tcolorbox}
\end{center}
}
\begin{document}

\title{\bf On the Popularity of GitHub Applications: \\ A Preliminary Note}


\author{Hudson Borges, Marco Tulio Valente, Andre Hora, Jailton Coelho \\[0cm] 
	{\normalsize Department of Computer Science, UFMG, Brazil} \\[0cm]  
	{\small \{hsborges,mtov,hora,jailtoncoelho\}@dcc.ufmg.br} 
}

\date{}

\maketitle

\begin{abstract}
\noindent GitHub is the world's largest collection of open source software. Therefore, it is important both to software developers and users to compare and track the popularity of  GitHub repositories. In this paper, we propose a framework to assess the popularity of GitHub software. We also propose a set of popularity growth patterns, which describe the evolution of the number of stars of a system over time. We show that stars tend to correlate with other measures, like forks, and with the effective usage of GitHub software by third-party programs. Throughout the paper we illustrate the application of our framework using real data extracted from GitHub.
\end{abstract}

\section{Introduction}

The popularity of a software is a valuable information both to its developers and to its users~\cite{Mileva12}. On one hand, producers constantly want to know if their systems are attracting new users, if the new releases are gaining acceptance, if they are meeting the users' expectations, etc.  On the other hand, users typically want to know if others are indeed sharing their decision of using a given system, if a system is mature enough for being used, if it is facing the risk of discontinuation, etc. 

In this paper, we describe a framework for assessing the popularity of open source systems hosted at GitHub, which is nowadays the largest source code sharing platform in the world, with around 10M users and 24M repositories.\footnote{https://github.com/about/press, verified on 06/17/2015.}  Interestingly, GitHub provides an explicit feature for users to manifest their interest or satisfaction with a hosted repository: the stargazers button. Like in social websites, users can {\em like} or {\em upvote} a repository by clicking in this button. Therefore, we assume in this paper that the number of stars of a repository is a reliable and simple measure of its popularity in GitHub ecosystem. We then make four contributions: (a) we propose a criteria to classify popular and very popular systems at GitHub (Section \ref{sec:popularity}); (b) we propose four growth patterns to describe the evolution of the number of stars of a system over time (Section~\ref{sec:growth-patterns}); (c) we verify the importance of the stargazers measure, by correlating the number of stars with other measures such as forks and client usage (Section~\ref{sec:correlations}). Throughout of the paper, we illustrate the application of our framework on a snapshot of GitHub, collected on May, 1st, 2015.



\section{Measuring Popularity}
\label{sec:popularity}

\begin{figure}[!h]
\centering
\vspace{-.2cm}
\includegraphics[width=0.65\linewidth]{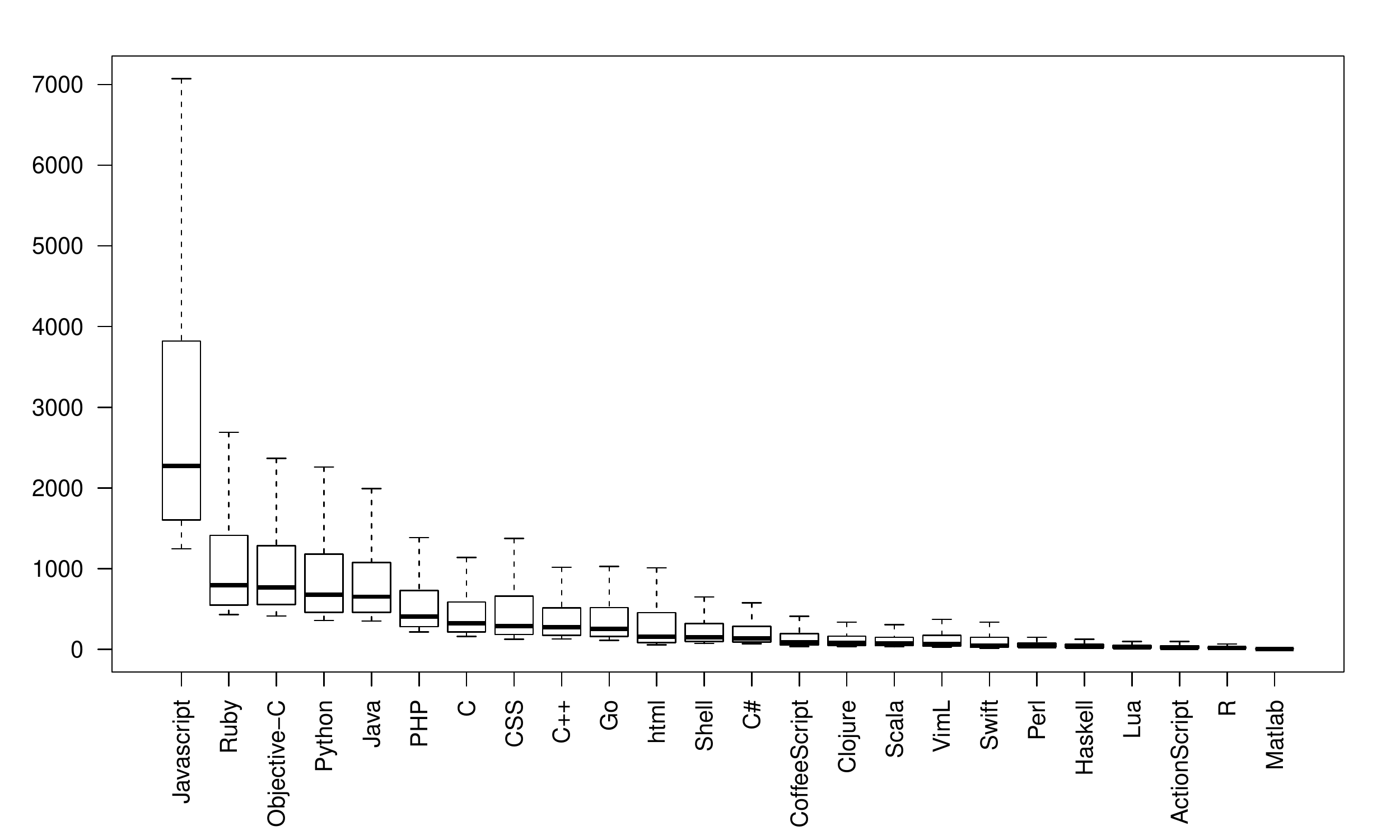}
\caption{Stars by programming language (top-1,000 systems, from top-24 popular languages)}
\label{fig:languages-overview}
\end{figure}

We consider the top-24 programming languages with more repositories in GitHub, which are the languages classified as popular by the GitHub advanced search engine\footnote{https://github.com/search/advanced}. Figure~\ref{fig:languages-overview} shows the distribution of the number of stars in the top-1,000 repositories of such languages. JavaScript is the language with the highest number of popular systems; in our sample of 1,000 JavaScript systems, the first quartile (bottom-25\%), second quartile (median), and third quartile (top-25\%) values are: 1,603, 2,274, and 3,820 stars. The median number of stars of the next five languages are: Ruby (793 stars), Objective-C (766 stars), Python (676 stars), Java (653 stars), and PHP (403 stars). The five languages with the lowest median number of stars are: Haskell (31 stars), Lua (24 stars), ActionScript (20), R (16 stars), and MatLab (4 stars).

We consider {\bf popular} the top-10\% systems with more stars in our sample of 24,000 systems, which requires a system to have at least 1,459 stars (on May, 1st, 2015). Table~\ref{tab:popular-repositories-by-language} shows the number of popular systems on each language. Furthermore, we call {\bf very popular} the top-1\% systems with more stars, which requires a system to have at least 19,570 stars.  Table~\ref{tab:popular-repositories-by-language} also shows the number of very popular systems, per programming language. Among these systems, 13 systems are implemented in JavaScript. Therefore, 35\% of the popular systems are in JavaScript, while the language has 54\% of the very popular ones. At the time of our analysis, {\sc twbs/bootstrap} is the most popular repository, followed by {\sc angular/angular.js}, and {\sc mbostock/d3}.

\begin{table}[!h]
\caption{Popular (top-10\%) and very popular (top-1\%) systems}
\centering
\begin{tabular}{lrr}
\toprule
 {\bf Language} & {\bf Popular} & {\bf Very Popular}\\
\midrule
Javascript	&	844	&	13	\\
Ruby	&	232	&	3	\\
Objective-C	&	207	&	0	\\
Python	&	194	&	0	\\
Java	&	159	&	0	\\
PHP	&	111	&	0	\\
CSS	&	108	&	3	\\
html	&	93	&	1	\\
Go	&	87	&	1	\\
C	&	79	&	1	\\
C++	&	69	&	1	\\
Shell	&	46	&	1	\\
CoffeeScript	&	42	&	0	\\
VimL	&	32	&	0	\\
C\#	&	25	&	0	\\
Scala	&	24	&	0	\\
Swift	&	22	&	0	\\
Clojure	&	10	&	0	\\
Haskell	&	6	&	0	\\
Perl	&	4	&	0	\\
ActionScript	&	3	&	0	\\
R	&	1	&	0	\\
Lua	&	1	&	0	\\
Matlab	&	1	&	0	\\
 \bottomrule  
Total	&	2,400	&	24	\\
 \bottomrule                                                                           
\end{tabular}
\label{tab:popular-repositories-by-language}
\end{table}

\section{Popularity Growth Patterns}
\label{sec:growth-patterns}

\begin{figure}[!h]
\centering
\vspace{-.7cm}
\includegraphics[width=0.6\linewidth]{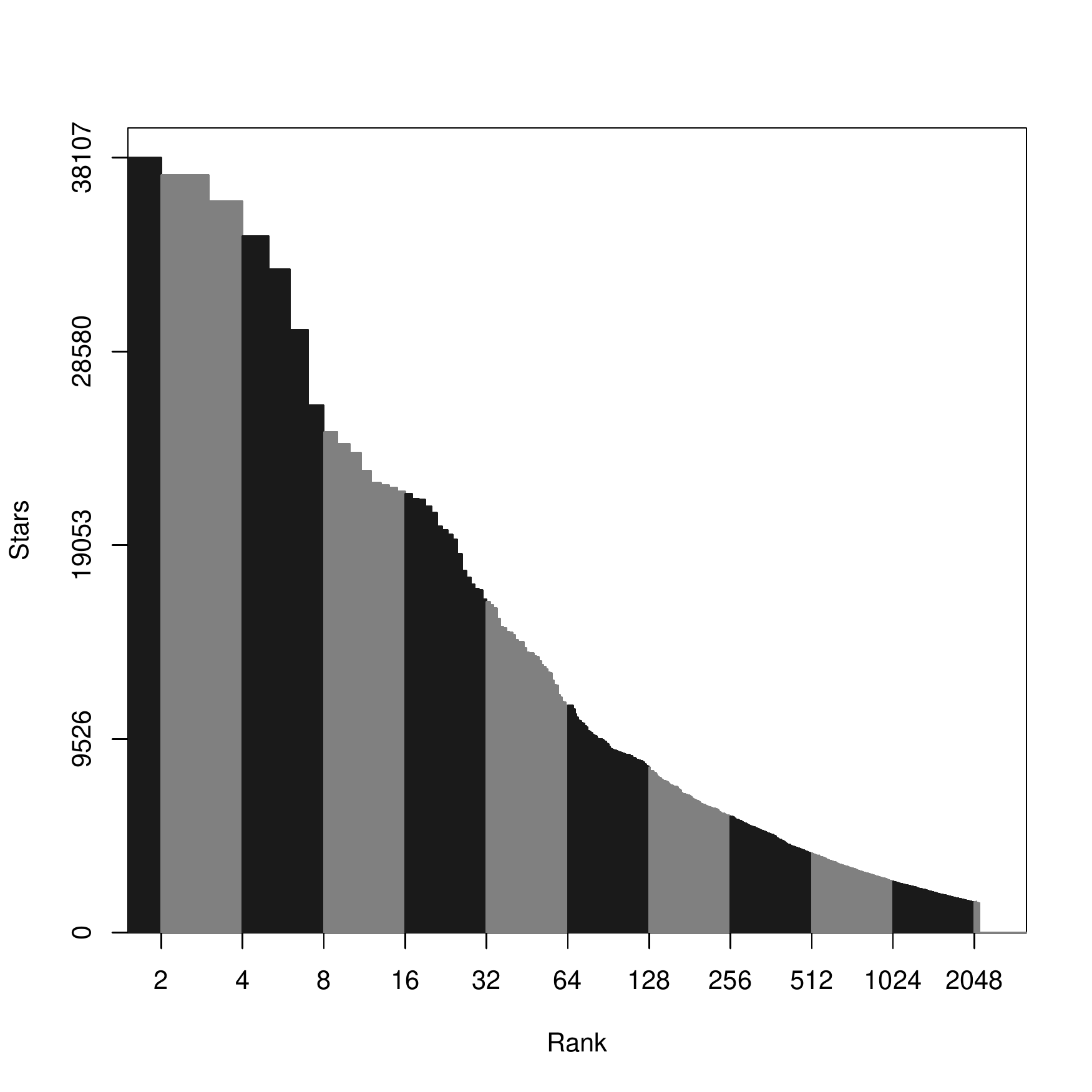}
\caption{Number of stars per rank position}
\label{fig:stars-distribution}
\end{figure}

For evaluating popularity over time, we restrict the analysis to popular systems with at least 52 weeks (one year) in order to include reasonable historical data for evaluation. We also excluded {\sc twbs/bootstrap} because we could not get all stars obtained by this system in the last 52 weeks. In this way, we study 2,138 popular systems (89\% of our initial set of popular systems). For  a given system, we define that $R_t$ is its rank in our list of popular systems in the week $t$ in a logarithm scale (base 2). Therefore, $log_2(1)+1 \leq R_t \leq log_2 (2,138)+1$ and $1 \leq t \leq 52$. The ranks are considered in logarithm scale due to the right-skewed distribution in the number of stars of the the popular systems, as presented in Figure~\ref{fig:stars-distribution}. The rank of the most popular system is 1. The earliest week is the week 1 (aka as $\mathit{OLD})$ and the latest one is 52 (aka as $\mathit{NEW})$. We also define that $R_{TOP}$ and $R_{BOTTOM}$ are respectively the highest (best) and lowest rank (worse) of a system in the interval under analysis.

\begin{figure}[!t]
\centering
\begin{subfigure}[t]{0.32\textwidth}
	\caption{{\sc Rails}}
	\includegraphics[width=\textwidth, trim=0cm 0cm 0cm 2cm, clip=true]{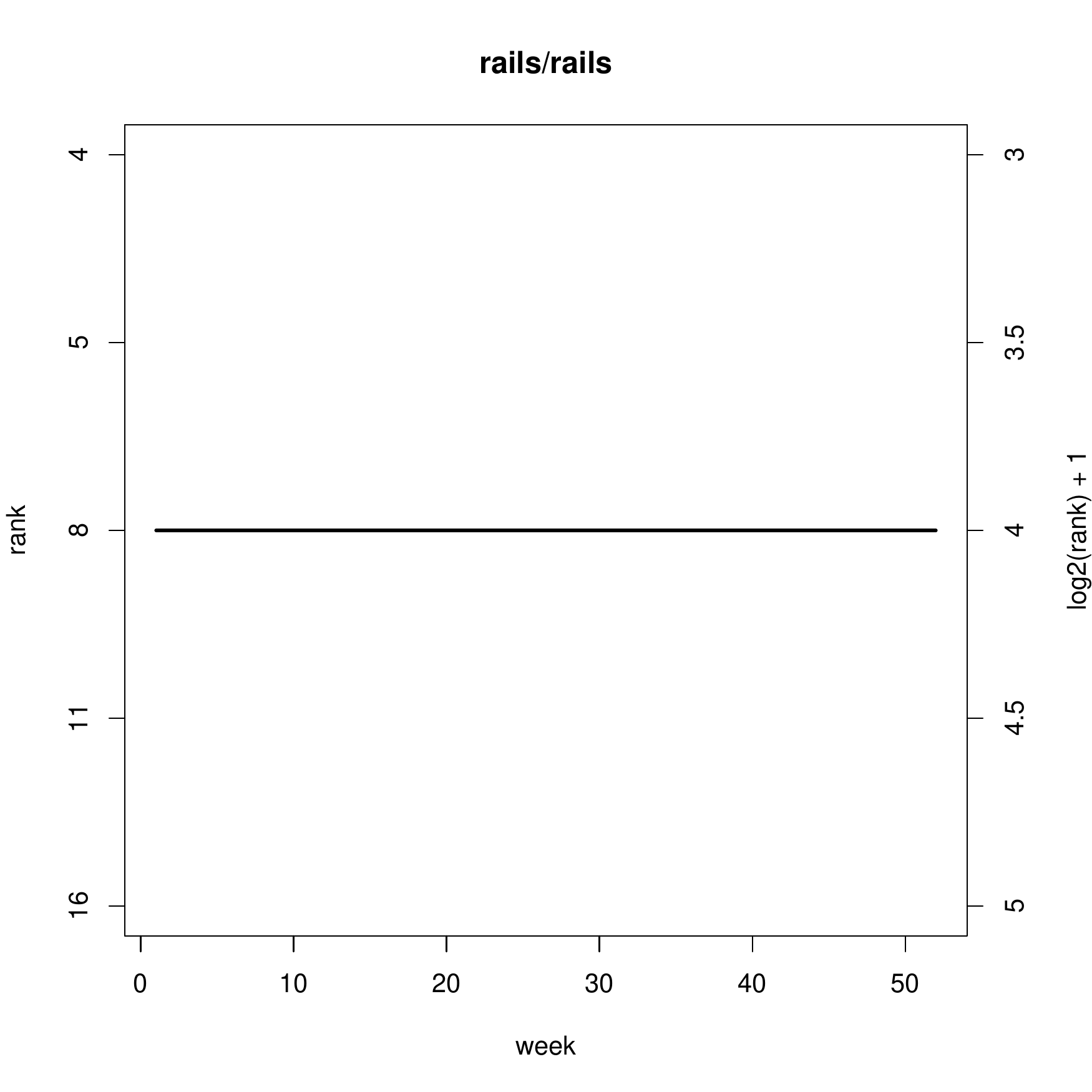}
	\vspace{-0.7cm}
\end{subfigure}
\begin{subfigure}[t]{0.32\textwidth}
 	\caption{{\sc Facebook Shimmer}}
	\includegraphics[width=\textwidth, trim=0cm 0cm 0cm 2cm, clip=true]{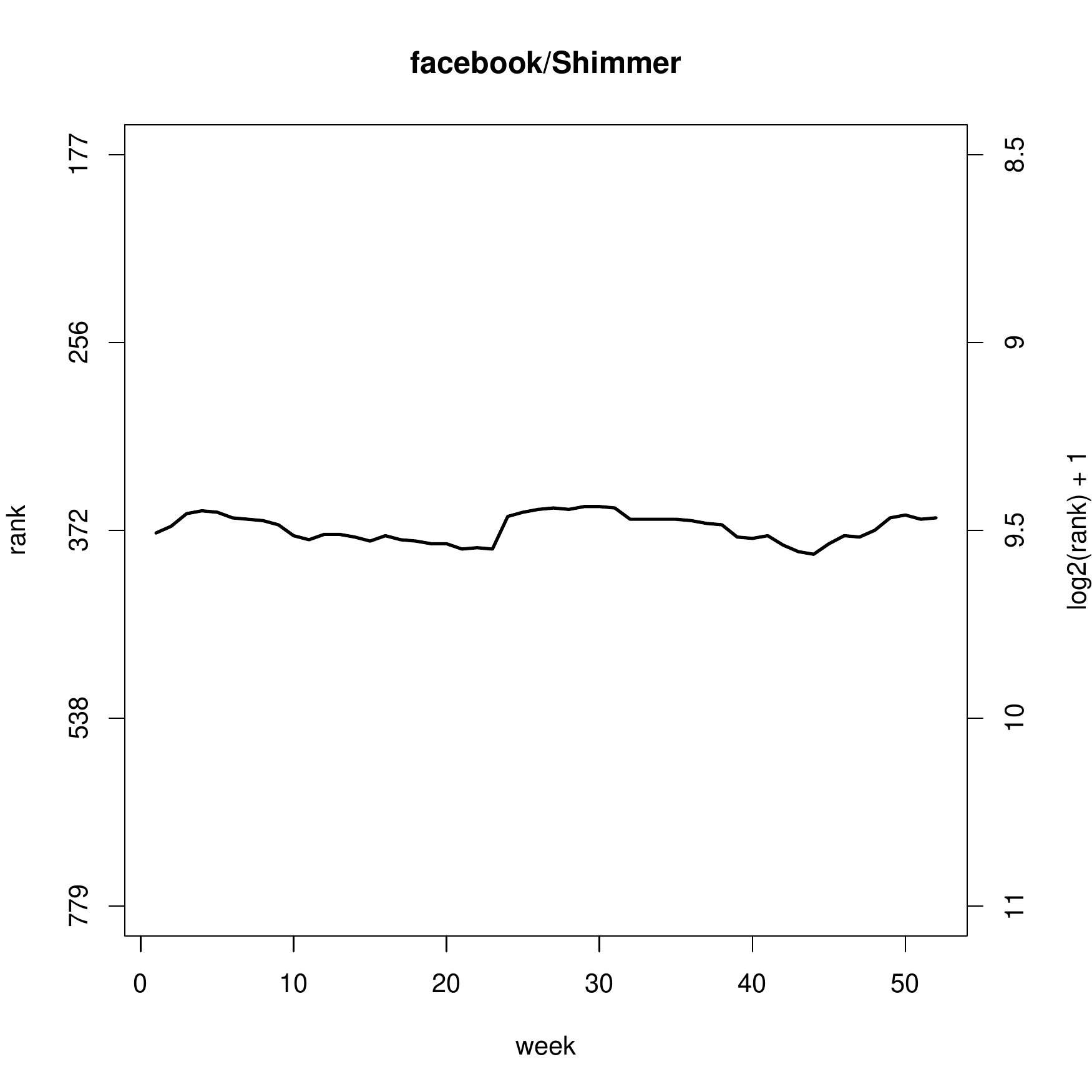}
	\vspace{-0.7cm}
\end{subfigure}
\begin{subfigure}[t]{0.32\textwidth}
	\caption{{\sc Socket.io}}
	\includegraphics[width=\textwidth, trim=0cm 0cm 0cm 2cm, clip=true]{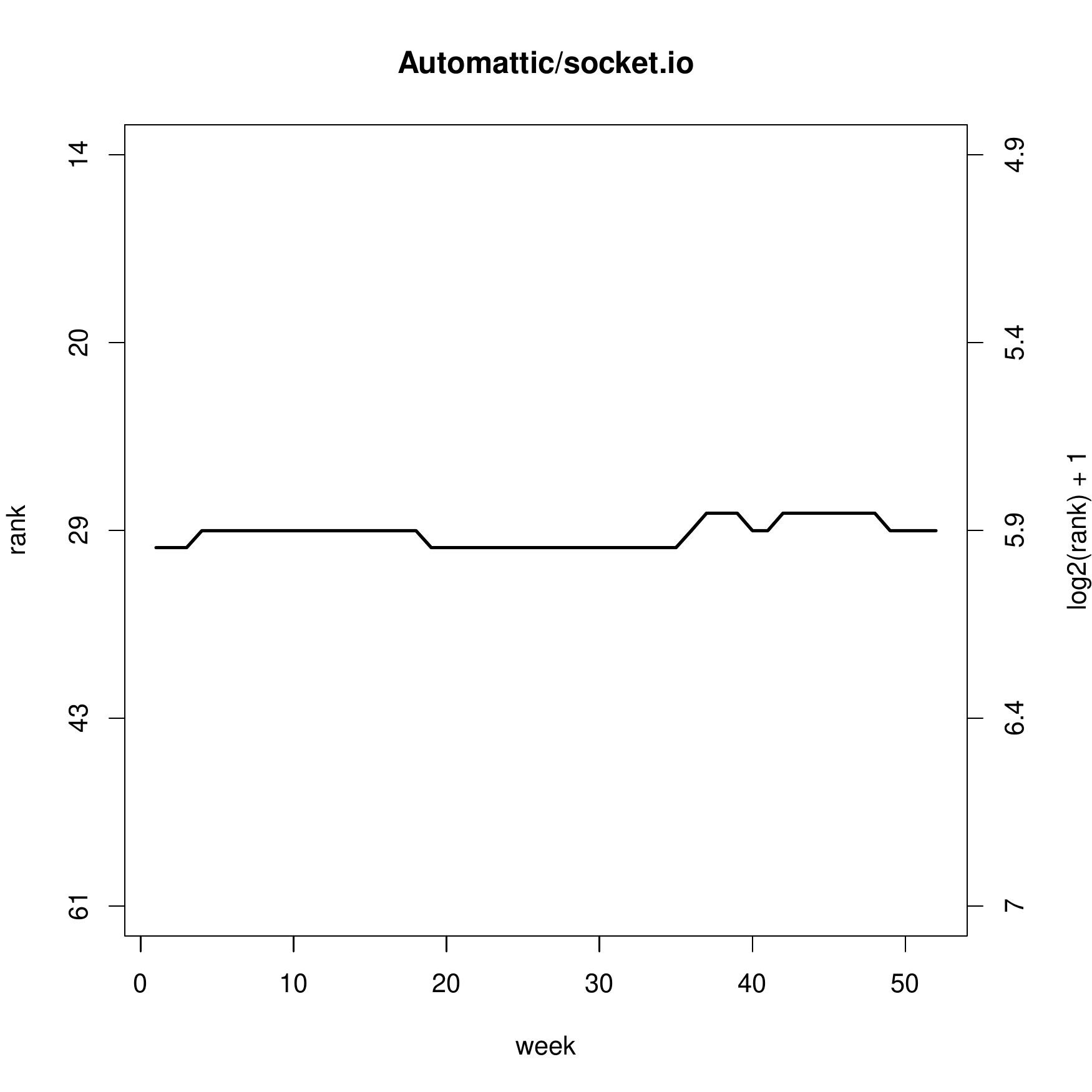}
	\vspace{-0.7cm}
\end{subfigure}
\linebreak
\vspace{-0cm}
\begin{subfigure}[t]{0.32\textwidth}
	\caption{{\sc Angular.js}}
	\includegraphics[width=\textwidth, trim=0cm 0cm 0cm 2cm, clip=true]{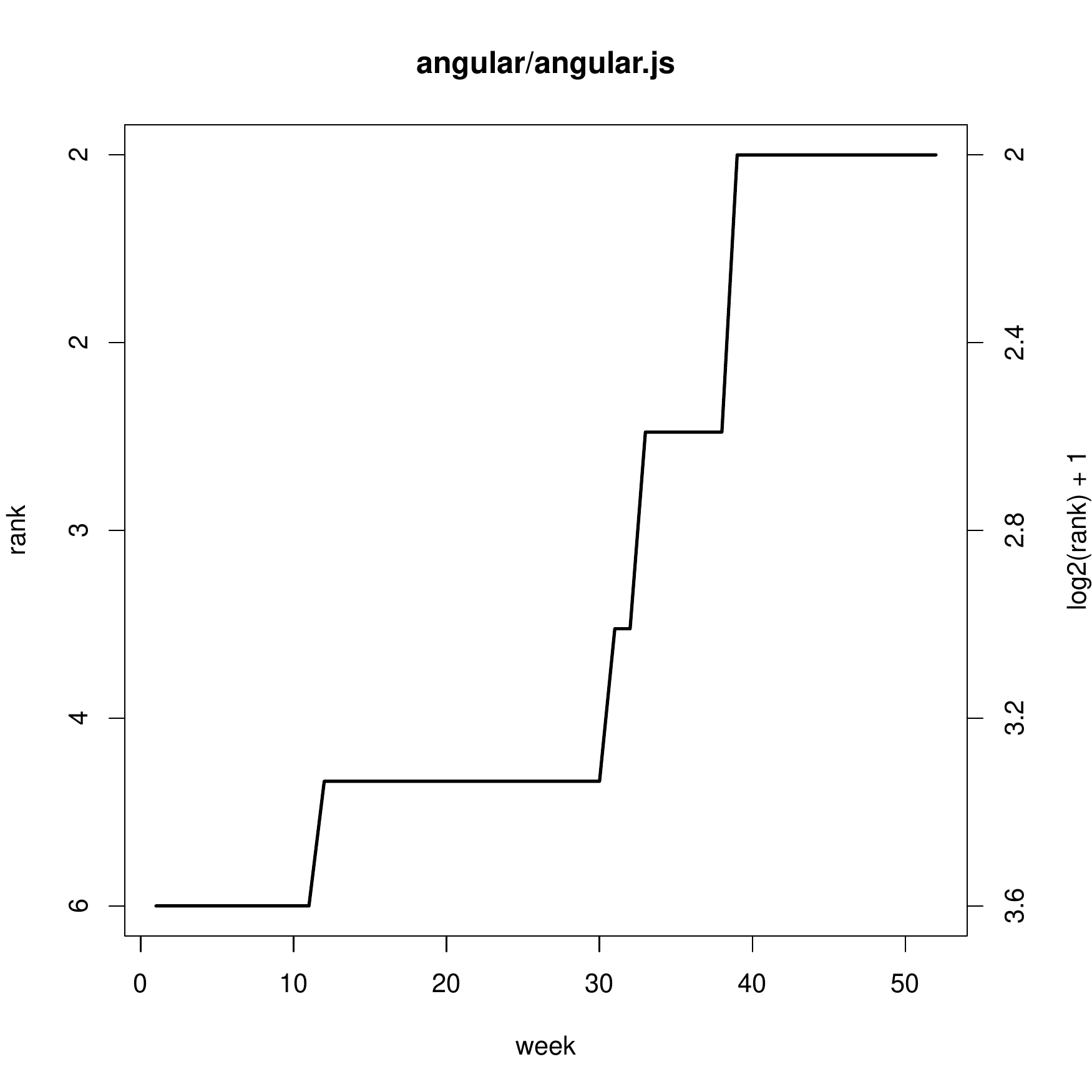}
	\vspace{-0.7cm}
\end{subfigure}
\begin{subfigure}[t]{0.32\textwidth}
	\caption{{\sc Docker Compose}}
	\includegraphics[width=\textwidth, trim=0cm 0cm 0cm 2cm, clip=true]{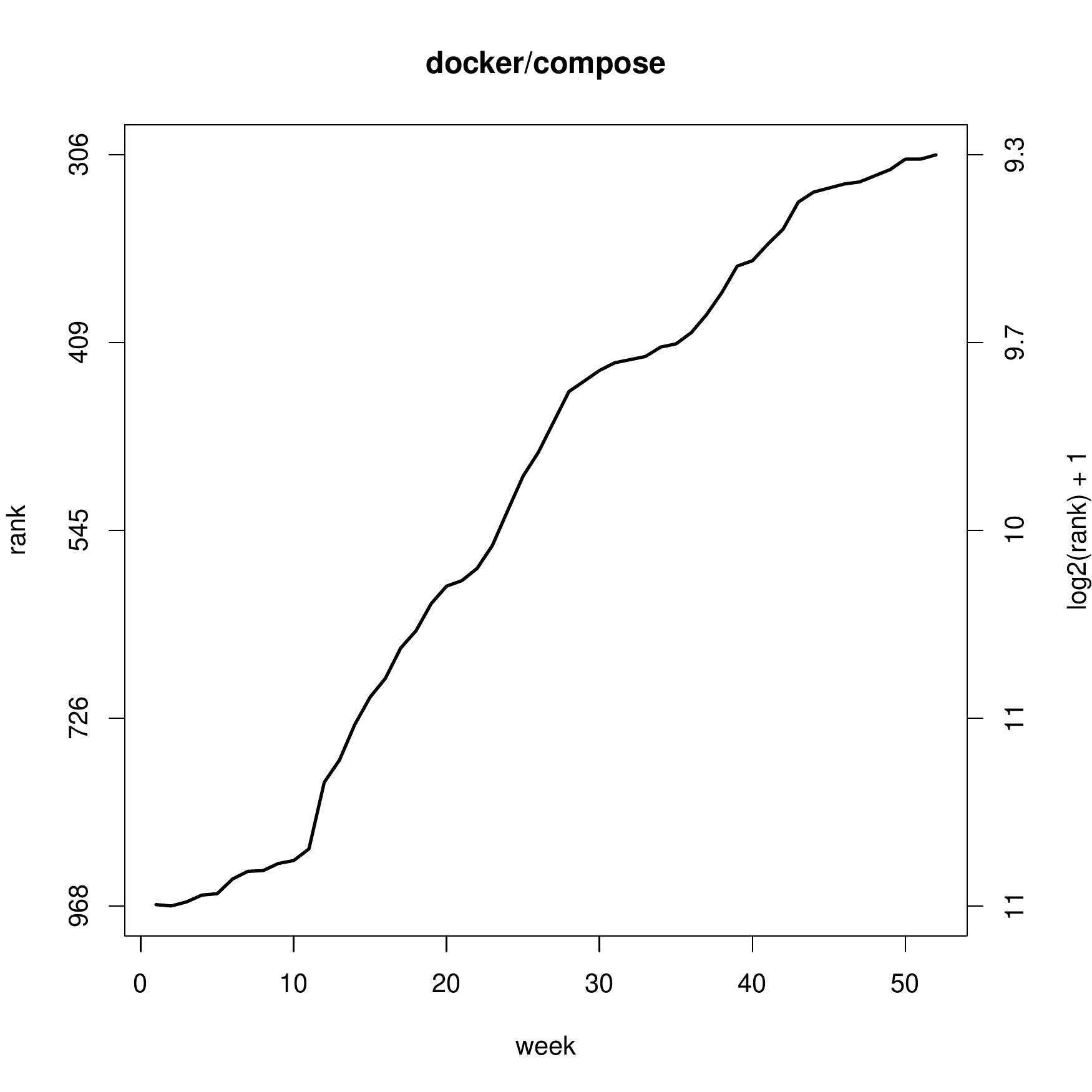}
	\vspace{-0.7cm}
\end{subfigure}
\begin{subfigure}[t]{0.32\textwidth}
	\caption{{\sc Apache Spark}}
	\includegraphics[width=\textwidth, trim=0cm 0cm 0cm 2cm, clip=true]{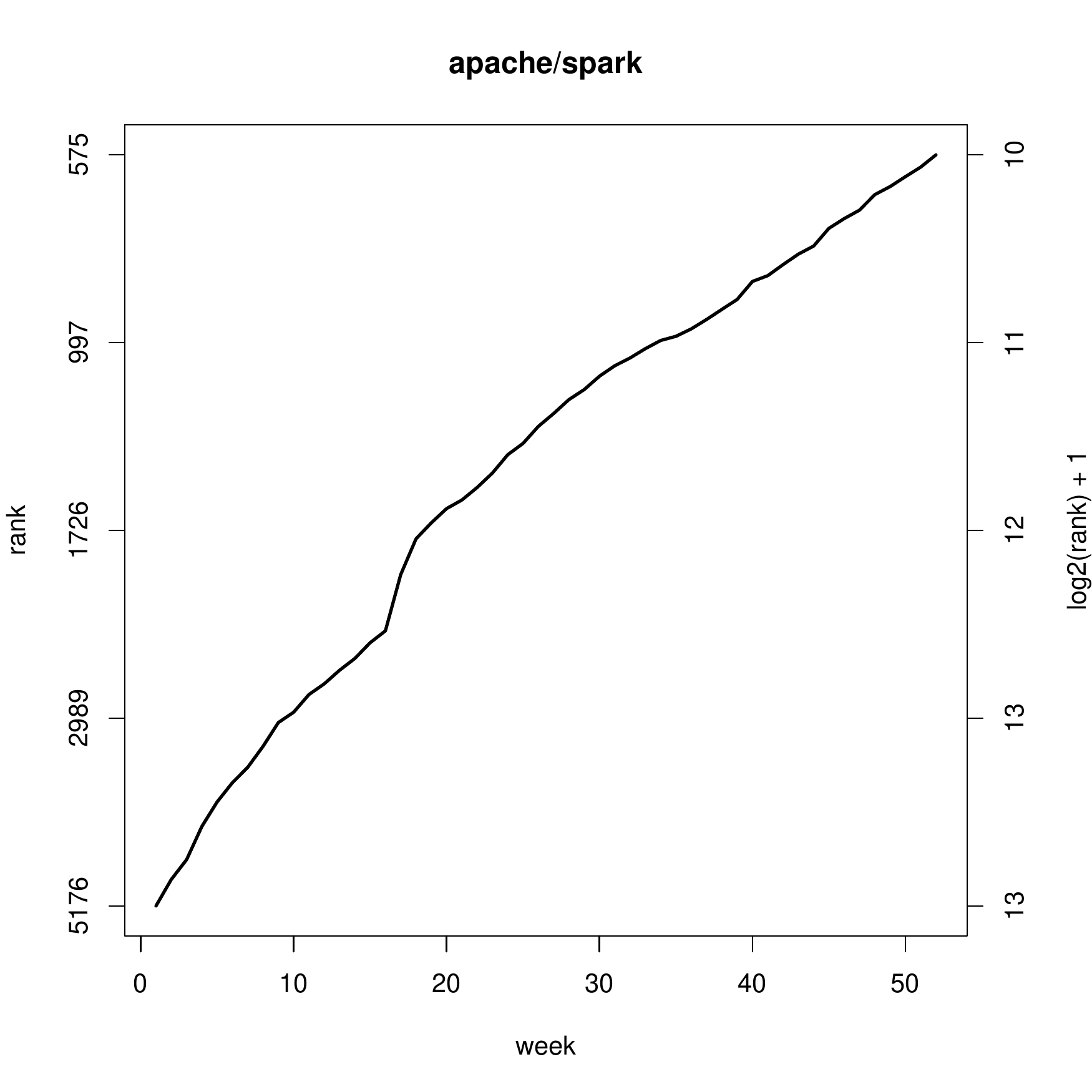}
	\vspace{-0.7cm}
\end{subfigure}
\linebreak
\vspace{-0cm}
\begin{subfigure}[t]{0.32\textwidth}
	\caption{{\sc jQuery}}
	\includegraphics[width=\textwidth, trim=0cm 0cm 0cm 2cm, clip=true]{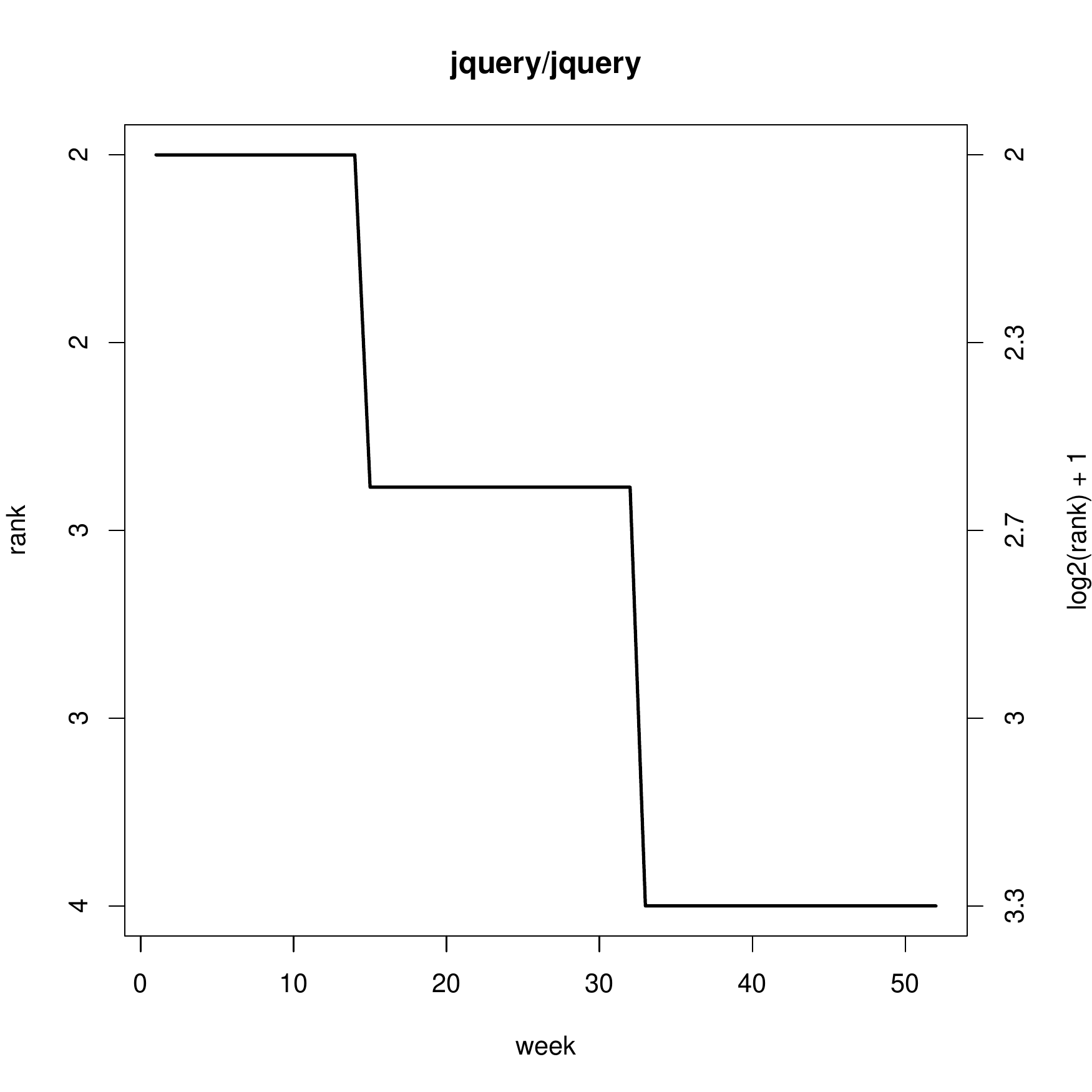}
	\vspace{-0.7cm}
\end{subfigure}
\begin{subfigure}[t]{0.32\textwidth}
	\caption{{\sc jQuery UI Bootstrap}}
	\includegraphics[width=\textwidth, trim=0cm 0cm 0cm 2cm, clip=true]{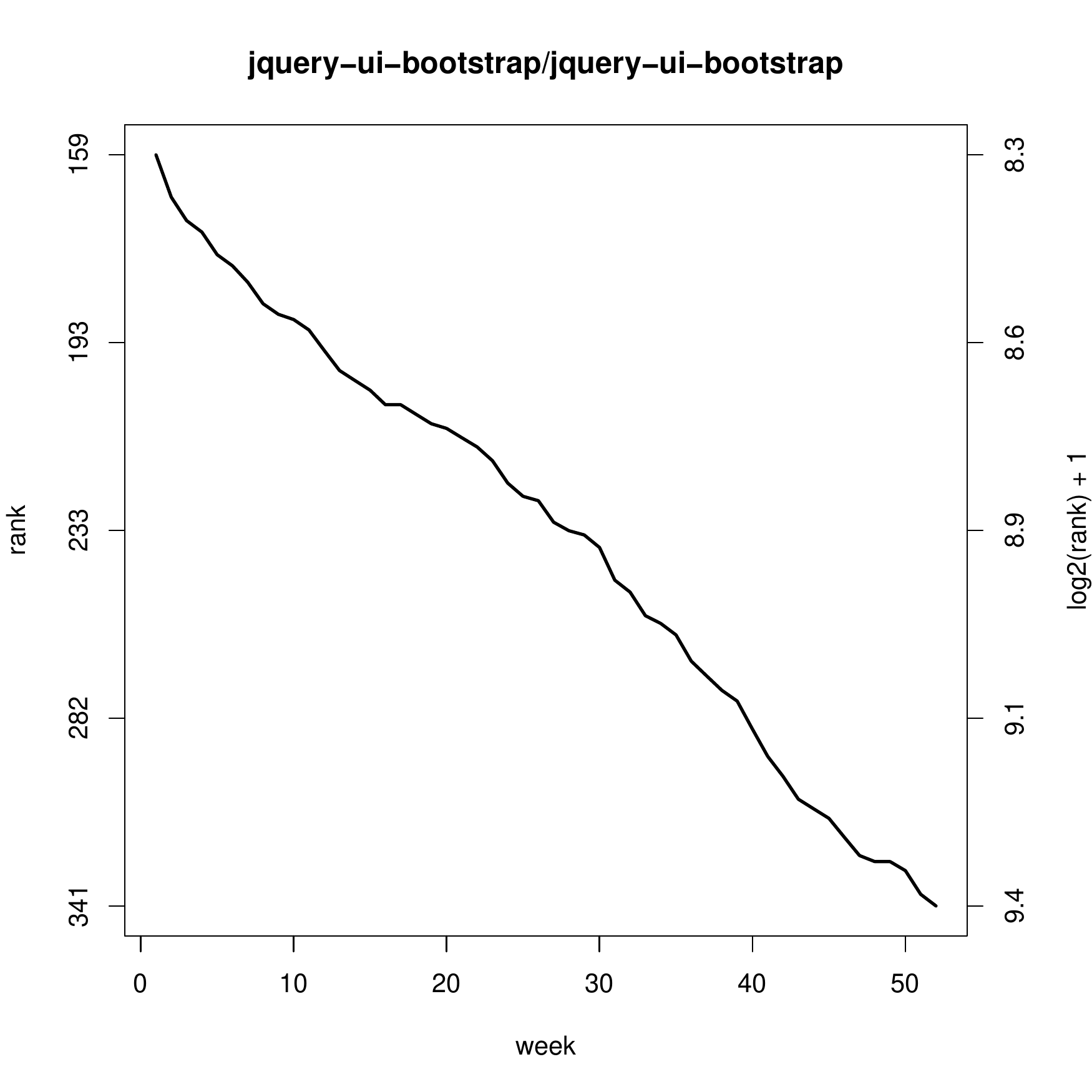}
	\vspace{-0.7cm}
\end{subfigure}
\begin{subfigure}[t]{0.32\textwidth}
	\caption{{\sc Django Old}}
	\includegraphics[width=\textwidth, trim=0cm 0cm 0cm 2cm, clip=true]{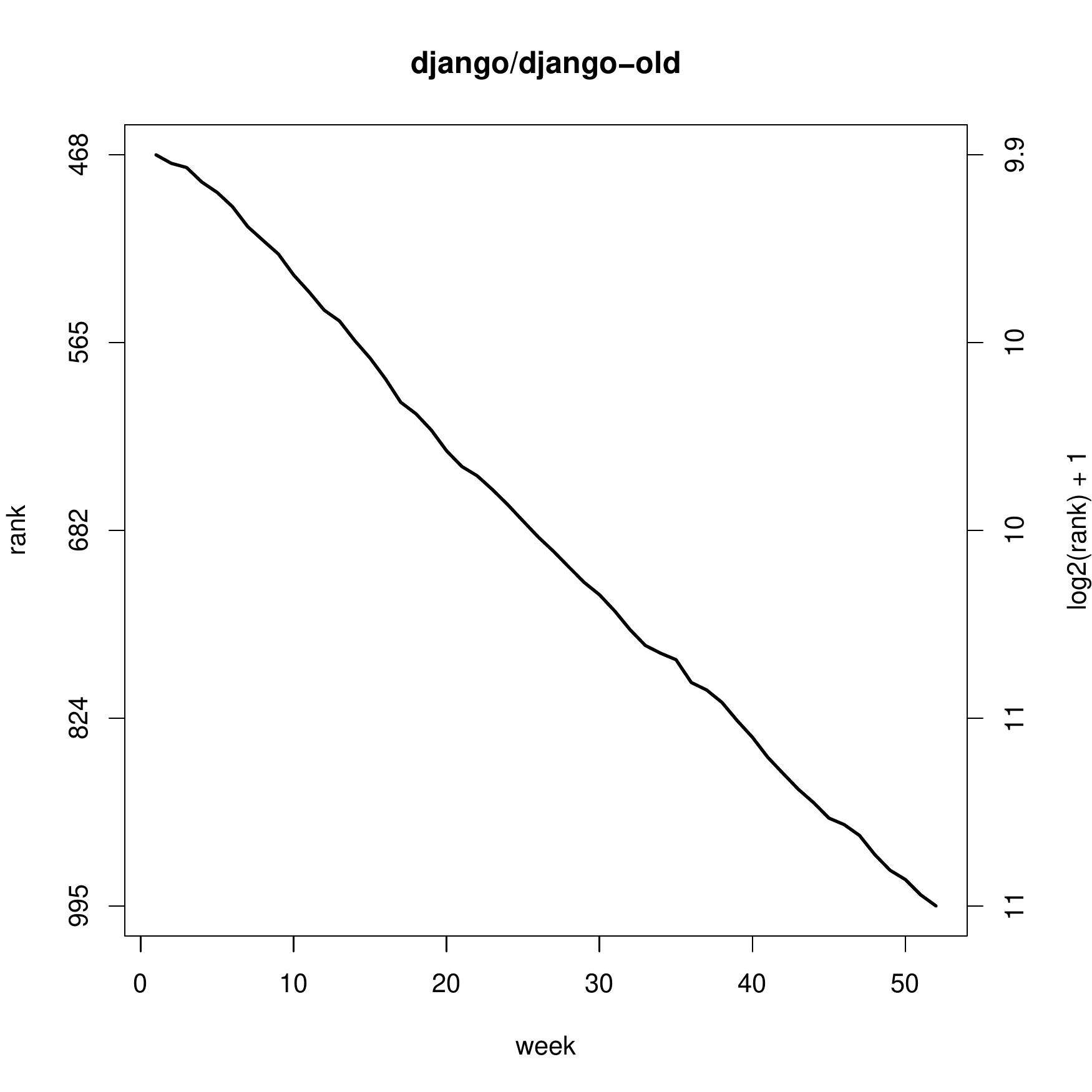}
	\vspace{-0.7cm}
\end{subfigure}
\linebreak
\vspace{-0cm}
\begin{subfigure}[t]{0.32\textwidth}
	\caption{{\sc KaTeX}}
	\vspace{0.475cm}
	\includegraphics[width=\textwidth, trim=0cm 0cm 0cm 2cm, clip=true]{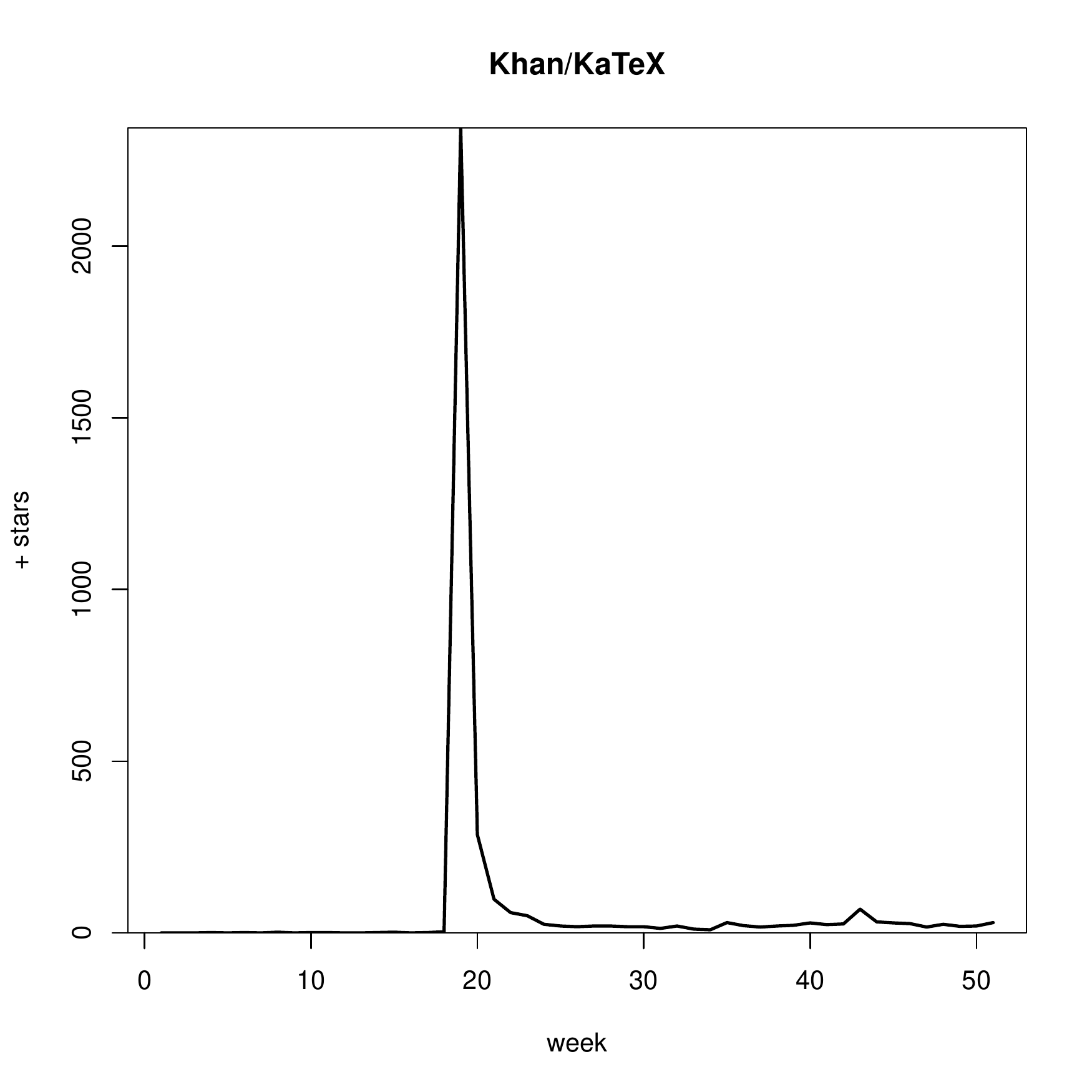}
\end{subfigure}
\begin{subfigure}[t]{0.32\textwidth}
	\caption{{\sc Augmented Traffic Control}}
	\includegraphics[width=\textwidth, trim=0cm 0cm 0cm 2cm, clip=true]{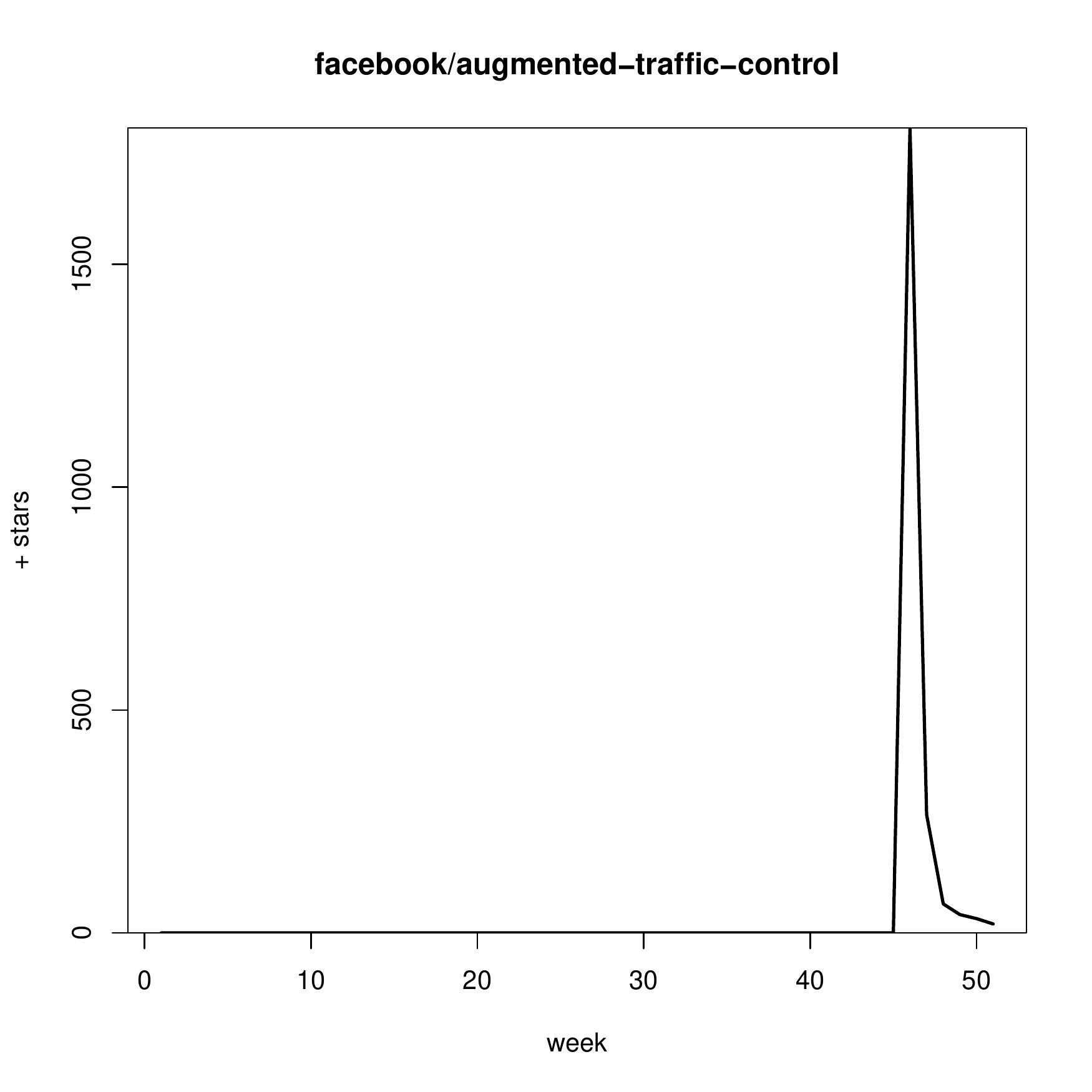}
\end{subfigure}
\begin{subfigure}[t]{0.32\textwidth}
	\caption{{\sc Git Large File Storage}}
	\vspace{0.475cm}
	\includegraphics[width=\textwidth, trim=0cm 0cm 0cm 2cm, clip=true]{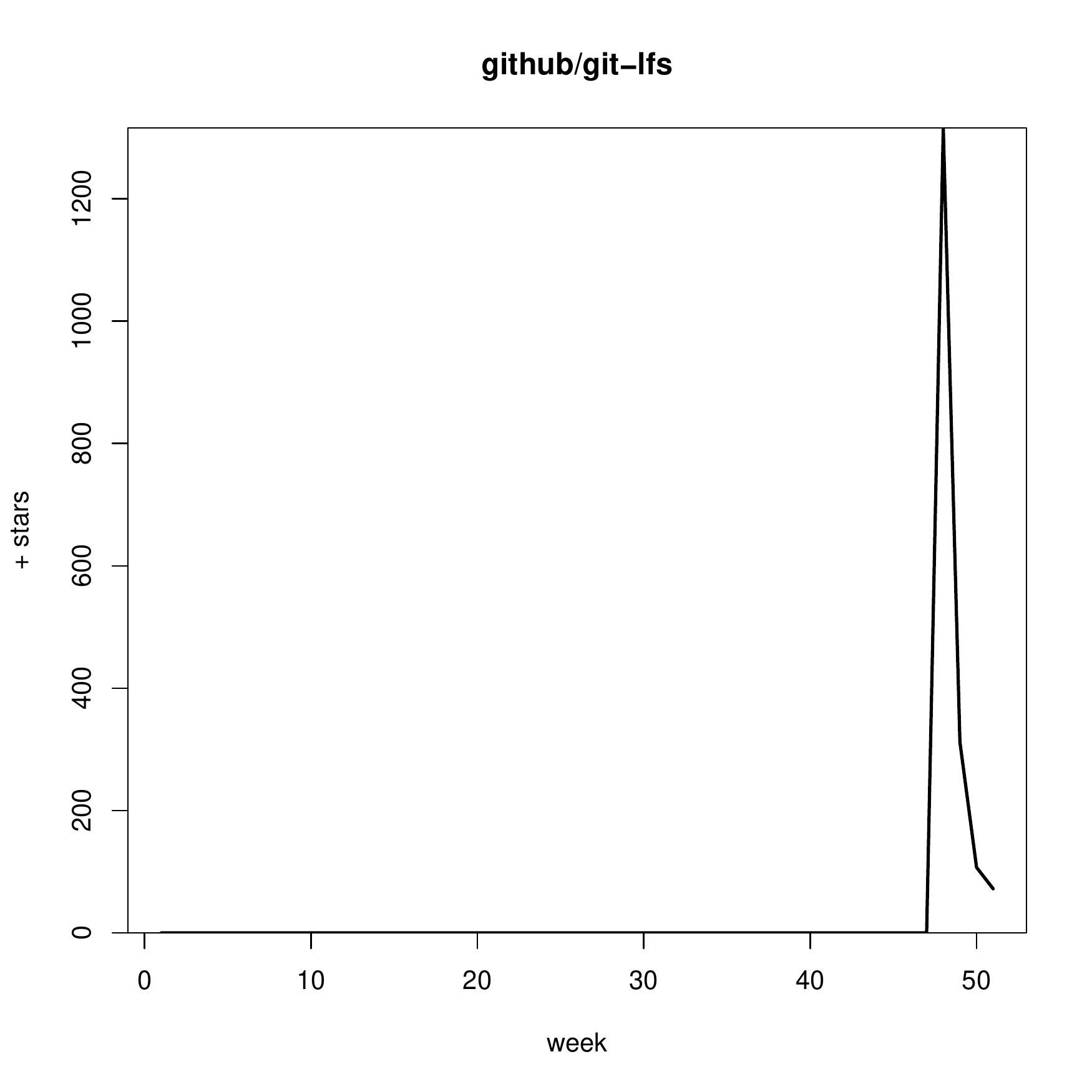}
\end{subfigure}
\caption{Growth Patterns: sustainable (a to c), fast (d to f), slow (g to i), and viral (j to l)}
\label{fig:growth}
\end{figure}

We propose four patterns of popularity growth: Sustainable, Fast, Slow, and Viral. These patterns are described next:\\

\noindent {\bf Sustainable Growth:} These systems sustained their ranking in the period under analysis, \emph{i.e.,} the number of stars they received in an one-year interval was sufficient to preserve their position in the ranking of popular systems. We use the following relation to express a sustainable growth:
\[(R_{BOTTOM} - R_{TOP}) < 0.25\]
In other words, a system with sustainable growth has minor variations in its rank during the period under analysis, inferior to 0.25 points in a logarithmic scale. Table~\ref{tab:regression} shows the number of systems with a sustainable growth per language. As can be observed, 466 systems (22\%) matched our definition for sustainable growth. Figures~\ref{fig:growth}a to \ref{fig:growth}c show three examples of such systems: {\sc rails/rails}, {\sc facebook/Shimmer}, and {\sc Automattic/socket.io}. \\


\noindent {\bf Fast Growth:} These are trending systems, which gained stars in a quantity that resulted in a relevant increase in their position in the ranking of popular systems, as captured by the following relation:
\[(R_{OLD} - R_{NOW}) > 1\,\,\    \wedge\,\,\   (R_{t+1} \leq R_{t})\,\,\, \mbox{in at least 90\% of the weeks t}\]
That is to say, a system with fast growth has now a rank position that is expressively better than its position one year ago, as expressed by a difference of at least one point in a logarithmic scale. Moreover,  in at least 90\% of the weeks under analysis the system preserved or improved its position in the ranking, compared to the previous week.  Table~\ref{tab:regression} shows the number of such systems per language. As can be observed, 100 systems (5\%) matched our definition for fast growth, including  44 systems in JavaScript and 18 systems in Java. Figures~\ref{fig:growth}d to \ref{fig:growth}f show three examples of such systems: {\sc angular/angular.js}, {\sc docker/compose}, and {\sc apache/spark}.\\


\noindent {\bf Slow Growth:} These are systems receiving few stars on each week. As a result, they experienced a relevant decrease in their rank position, as represented by the following relation:
\[(R_{NOW} - R_{OLD}) > 1\,\,\    \wedge\,\,\   (R_{t+1} \geq R_{t})\,\,\, \mbox{in at least 90\% of the weeks t}\]
A system with slow growth has now a ranking that is one logarithmic degree greater than its ranking one year ago. Moreover, in at least 90\% of the weeks under analysis the system preserved or decreased its position in the ranking, compared to the previous week. Table~\ref{tab:regression} shows the number of systems with slow growth per language. As presented in this table, 12 systems (0.5\%) matched our definition for this pattern. Figures~\ref{fig:growth}g to \ref{fig:growth}i show three examples of such systems: {\sc jquery/jquery}, {\sc jquery-ui-bootstrap/jquery-ui-bootstrap}, and {\sc django/django-old}.

\clearpage

\begin{table}[!h]
\caption{Popular systems following the proposed growth patterns}
\centering
\begin{tabular}{lrrrrr}
\toprule
 {\bf Language} & {\bf Systems} & {\bf Sustainable} & {\bf Fast} & {\bf Slow}\ & {\bf Viral} \\
\midrule
Javascript		&	762		&	146	&	44	&	2	&	13	\\
Ruby			&	225		&	30	&	0	&	3	&	2	\\
Objective-C		&	187		&	59	&	5	&	2	&	0	\\
Python			&	175		&	39	&	5	&	1	&	10	\\
Java			&	134		&	34	&	18	&	0	&	0	\\
PHP				&	105		&	42	&	4	&	0	&	0	\\
CSS				&	94		&	7	&	0	&	3	&	1	\\
html			&	83		&	26	&	2	&	1	&	3	\\
Go				&	62		&	14	&	10	&	0	&	2	\\
C				&	65		&	17	&	1	&	0	&	1	\\
C++				&	61		&	9	&	4	&	0	&	2	\\
Shell			&	43		&	7	&	0	&	0	&	1	\\
CoffeeScript	&	40		&	6	&	1	&	0	&	0	\\
VimL			&	32		&	12	&	2	&	0	&	0	\\
C\#				&	20		&	5	&	3	&	0	&	0	\\
Scala			&	24		&	7	&	1	&	0	&	0	\\
Swift			&	1		&	0	&	0	&	0	&	1	\\
Clojure			&	10		&	3	&	0	&	0	&	0	\\
Haskell			&	5		&	1	&	0	&	0	&	1	\\
Perl			&	4		&	1	&	0	&	0	&	0	\\
ActionScript	&	3		&	0	&	0	&	0	&	0	\\
R				&	1		&	1	&	0	&	0	&	0	\\
Lua				&	1		&	0	&	0	&	0	&	0	\\
Matlab			&	1		&	0	&	0	&	0	&	0	\\
 \bottomrule         
Total			&	2,138	&	466	&	100	&	12	&	37	\\
 \bottomrule                                                                           
\end{tabular}
\label{tab:regression}
\end{table}

\noindent {\bf Viral Growth:}  These systems experience a  massive growth in their number of stars in a short period of time. Typically, viral growth results from word-of-mouth propagation in 
social networks (Twitter, Facebook, etc) or social news sites (Hacker News, Slashdot, Reddit, etc). In this paper, we consider that viral growth happens when a systems receives most of its stars (\emph{i.e.,}~$>$ 50\%) in a single week.  Table~\ref{tab:regression} also shows the number of popular systems classified as viral. We found 37 viral systems (2\%) and JavaScript is the language with the highest number of such systems (13 systems). Figures~\ref{fig:growth}j to~\ref{fig:growth}k show three examples of systems with a viral growth. We can see that they received more than 1,300 stars in a single week, and very few stars (if any) in the other weeks. To illustrate the importance of social sites on viral growth, we checked that a post on {\sc Khan/KaTeX} (a JavaScript library for \TeX\ math rendering) was heavily commented and upvoted at Hacker News in the week the system experienced the massive peak in the number of stars\footnote{https://news.ycombinator.com/item?id=8320439}. {\sc facebook/augmented-traffic-control} (a network connection simulator) and {\sc github/git-lfs} (a tool for managing large files with Git) are other systems that attracted media coverage exactly in the week they received more than 1,800 stars\footnote{http://www.wired.com/2015/03/facebook-traffic-control} and 1,300 stars\footnote{https://github.com/blog/1986-announcing-git-large-file-storage-lfs}, respectively.

\section{Correlating Popularity with Forks and Usage}
\label{sec:correlations}

To clarify the credibility of the number of stars as measure for a system's popularity, we investigate the correlation between this measure and two other ones: number of forks and number of clients.\\

\begin{figure*}[!t]
\centering

\begin{subfigure}[t]{0.49\textwidth}
	\includegraphics[width=\textwidth]{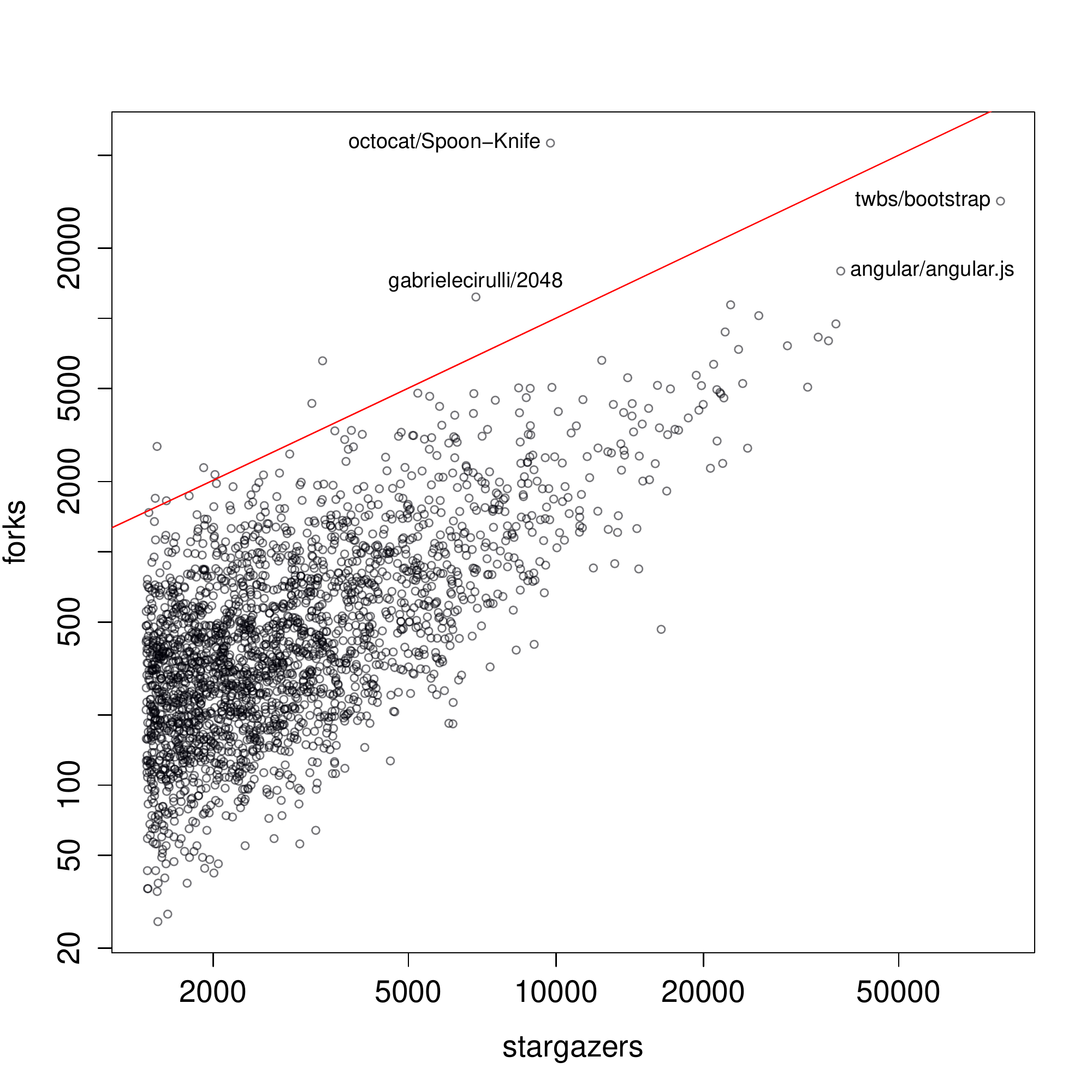}
	\vspace{-0.8cm}
	\caption{Popularity  vs number of forks}
	\label{fig:correlations:forks}
\end{subfigure}
\begin{subfigure}[t]{0.49\textwidth}
	\includegraphics[width=1\textwidth]{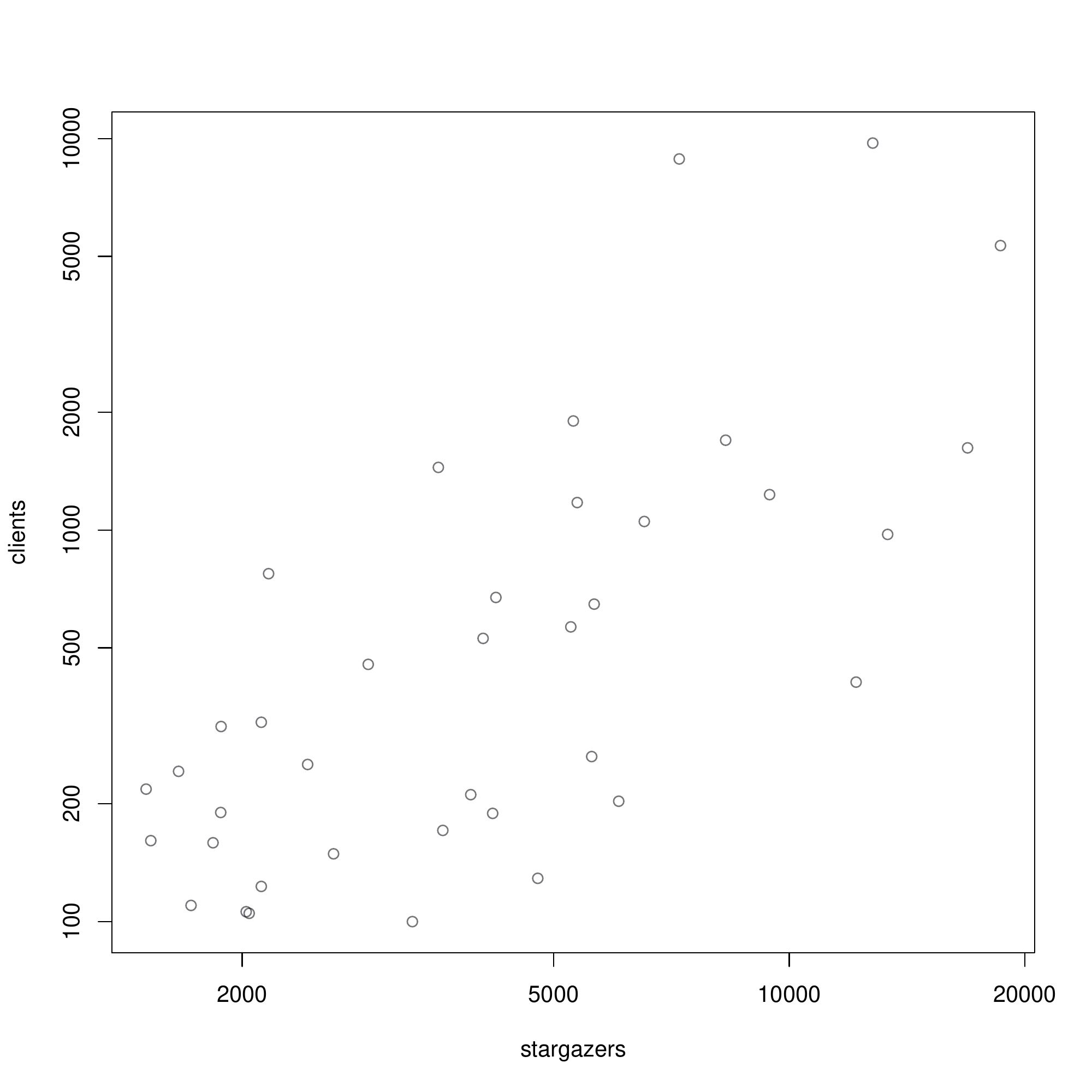}
	\vspace{-0.8cm}
	\caption{Popularity  vs clients (for 38 Node.js apps)}
	\label{fig:correlations:clients}
\end{subfigure}

\caption{Popularity correlation with forks and usage}
\label{fig:correlations}
\end{figure*}

\noindent{\em Forks: } In git-based systems, forks are used to either propose changes to an application or as a starting point for a new project. In both cases, the number of forks can be seen as a proxy for the importance of a project in GitHub. Figure~\ref{fig:correlations:forks} shows plots correlating a system popularity and its number of forks. A logarithm scale is used in both axes. The line represents the identity relation: below the line are the systems with more stars than forks, and above the line the opposite. Two facts can be observed in this figure. First, there is a strong positive correlation between stars and forks (Spearman rank correlation coefficient = 0.55). Second, only a few systems have more forks than stars. 
As examples, we have a repository that just provides an example for forking a repository on GitHub ({\sc octocat/Spoon−Knife}) and a popular puzzle game ({\sc gabrielecirulli/2048}), whose success motivated many forks with variations of the original implementation. Since the game can be downloaded directly from the web, we hypothesize that it receives most users's feedback in the web and not on GitHub.\\

\noindent{\em Clients:} The number of clients is another clear measure of popularity. However, it is not trivial to access the clients of most GitHub applications. For example, {\sc mbostock/d3} is a popular visualization library, which can be imported by any Web page, including public and private ones. Therefore, it is not trivial to search for D3's clients. For this reason, to correlate client usage and stars, we focus on a restricted set of applications, composed by Node.js-based libraries hosted on the NPM registry. Node.js is a popular, runtime environment for server-side and networking JavaScript applications. NPM (Node Package Manager) is a centralized repository for hosting the production version of JavaScript modules. Although, NPM can host any JavaScript module, it is the {\em de facto} platform for hosting Node.js-based applications. Moreover, NPM's API provides means for accessing the number of clients---or dependents, in NPM terms---of a given module. Therefore, we first retrieved the number of dependents of the popular JavaScript applications in our dataset, using the NPM API. We then manually inspected the top-100 applications in terms of dependents to select the Node.js-based modules. We found 38 of such systems, which are listed in Table~\ref{tab:nodejs}. Figure~\ref{fig:correlations:clients} shows a plot correlating the values we found for number of stars and number of dependents. As can be visually observed, there is a strong correlation between these two measures, with a Spearman's rank correlation coefficient of 0.68.

\begin{table}[!h]
\footnotesize
\caption{Popular Node.js-based libraries}
\centering
\begin{tabular}{lllll}
\toprule

strongloop/express & 
Automattic/socket.io & 
gulpjs/gulp & 
caolan/async \\
bower/bower & 
gruntjs/grunt & 
jadejs/jade & 
request/request \\ 
mochajs/mocha & 
koajs/koa &
stylus/stylus & 
babel/babel \\ 
senchalabs/connect & 
cheeriojs/cheerio & 
jaredhanson/passport &
webpack/webpack \\ 
tmpvar/jsdom & 
hapijs/hapi & 
andris9/Nodemailer & 
twitter/hogan.js \\
sequelize/sequelize & 
winstonjs/winston & 
NaturalNode/natural & 
tj/co \\ 
louischatriot/nedb &
postcss/postcss & 
websockets/ws & 
aheckmann/gm \\ 
reworkcss/rework & 
substack/dnode &
olado/doT & 
paularmstrong/swig \\ 
creationix/step & 
mozilla/nunjucks & 
sstephenson/eco &
sass/node-sass \\ 
caolan/nodeunit & 
danwrong/restler \\

\bottomrule                                                                           
\end{tabular}
\label{tab:nodejs}
\end{table}

\sidebarcmd

\section{Conclusion}

We proposed a framework to track the popularity of GitHub systems and we found that: 

\begin{itemize}

\item JavaScript is responsible for more than one third of the popular applications on GitHub;  the next five languages (Ruby, Objective-C, Python, Java, and PHP) are responsible  for another third of the popular applications.

\item 21\% of the popular systems have a sustainable growth; 5\% have a fast growth; and less than 1\% have a slow growth. We also found 37 systems with a viral behavior.

\item The number of stars of a system tends to correlate not only with the number of forks, but also with its effective usage by other client applications, which reinforces the importance of stars as a real measure of a system's popularity.
\end{itemize}

A possible implementation of this framework could be useful both to API users and to API developers, who certainly share interest on monitoring the usage and popularity of their APIs over time.

\section*{Acknowledgment}

\noindent Our research is supported by CNPq and FAPEMIG.

\bibliographystyle{IEEEtran}
\bibliography{bibfile}

\end{document}